\documentclass[twocolumn,aps,prx,superscriptaddress]{revtex4}
\usepackage{graphicx}
\usepackage{amsmath,amssymb,bm}
\usepackage{braket}
\usepackage[colorlinks=true,linkcolor=blue,citecolor=blue,urlcolor=blue]{hyperref}

\begin{document}

\title{Noise Enhanced Neural Networks for Analytic Continuation}

\author{Juan Yao}
\affiliation{Shenzhen Institute for Quantum Science and Engineering, Southern University of Science and Technology, Shenzhen 518055, Guangdong, China}
\author{Ce Wang}
\affiliation{Institute for Advanced Study, Tsinghua University, Beijing, 100084, China}
\author{Zhiyuan Yao}
\affiliation{Institute for Advanced Study, Tsinghua University, Beijing, 100084, China}
\author{Hui Zhai}
\affiliation{Institute for Advanced Study, Tsinghua University, Beijing, 100084, China}

\date{\today}
\begin{abstract}
Analytic continuation maps imaginary-time Green's functions obtained by various theoretical/numerical methods to real-time response functions that can be directly compared with experiments. Analytic continuation is an important bridge between many-body theories and experiments but is also a challenging problem because such mappings are ill-conditioned. In this work, we develop a neural network-based method for this problem. The training data is generated either using synthetic Gaussian-type spectral functions or from exactly solvable models where the analytic continuation can be obtained analytically. Then, we applied the trained neural network to the testing data, either with synthetic noise or intrinsic noise in Monte Carlo simulations. We conclude that the best performance is always achieved when a proper amount of noise is added to the training data. Moreover, our method can successfully capture multi-peak structure in the resulting response function for the cases with the best performance. The method can be combined with Monte Carlo simulations to compare with experiments on real-time dynamics.  
\end{abstract}
\maketitle

\section{Introduction}

Analytic continuation is an important problem in computational quantum many-body physics \cite{Werner,Correl12}. 
For quantum many-body systems, a variety of methods, ranging from perturbation approximations to quantum Monte Carlo algorithms, are performed in imaginary time and only produce imaginary-time Green's function \cite{Suzuki, Gull}. 
Nevertheless, to compare with real-time responses measured in experiments, we need the response function 
or the closely related spectral function $\mathcal{A}(\Omega)$ from which the response function can be calculated. On the other hand, the spectral function and the Matsubara frequency representation of the imaginary-time Green's function, $\mathcal{G}(i\omega_n)$, are related by \cite{Mahan, Simons} 
\begin{equation} \label{Analytical_continuation}
	\mathcal{G}(i\omega_n)=\int \frac{\mathcal{A}(\Omega)}{i\omega_n-\Omega} \, d\Omega  \, . 
\end{equation}
Hence this relation is crucial to analytically continue the imaginary-time Green's function to real time that connects numerical results to experimental observations.

However, extracting $\mathcal{A}(\Omega)$ form the above relation, Eq. \eqref{Analytical_continuation}, is a difficult problem. Although the mapping from $\mathcal{A}(\Omega)$ to $\mathcal{G}(i\omega_n)$ is linear, the coefficients decrease and approach zero as $\Omega$ increases, making the inverse of such mapping ill-conditioned \cite{Ramm, Kabanikhin1, Kabanikhin2}. 
That is to say, a small noise embedded in $\mathcal{G}(i\omega_n)$ can be significantly amplified and lead to huge errors in the evaluated $\mathcal{A}(\Omega)$. Since $\mathcal{G}(i\omega_n)$ obtained by various numerical methods inevitably contains noise, thus obtained $\mathcal{A}(\Omega)$ always has large errors, which makes the comparison with experimental measurements unreliable. This is the intrinsic difficulty in the analytic continuation \cite{Ramm, Kabanikhin1, Kabanikhin2}. 

Many methods have been developed to address this problem. One of the most popular approaches is the maximum entropy (MaxEnt) method which aims to find the most probable $A(\Omega)$ that maximizes an entropy functional \cite{Jarrell, Tremblay}. However, the method is biased towards Gaussian type spectral functions, and sharp peaks or edges in the spectral function can be easily missed. Several stochastic methods have then been proposed to recover these singular features \cite{Sandvik98, Fuchs, Sandvik16, Mishchenko,Goulko}. Nevertheless, these stochastic methods are usually time-consuming \cite{HYoon}. 
Many of the existing methods also rely on prior physical information. For instance, relying on the conjecture of holographic duality and the insight from gravity physics, the analytic continuation has also been applied to study quantum critical dynamics \cite{Subir}.

The fast development of applying machine learning methods to physics problems offer a new route to address this problem. The neural network (NN) is a useful tool to express functional mapping that can also tolerant errors. Nevertheless, the neural-network-based methods have only been discussed in few recent works on the analytic continuation problem \cite{Millis, HYoon, LeiWang, Huang}. While these method offer new promise, lots of issues remain open. Especially, since the difficulty of the analytic continuation roots in the amplification of the noise, one basis question is whether we should add noise into the training data and, if so, what is the proper amplitude of the noise. To the best of our knowledge, this issue has not been well studied before. In this work, we systematically study the noise effect in the training dataset. The key finding is that the NN's performance can be significantly enhanced when a proper amount of noise is added to the data.

\begin{figure}[t!!]
\begin{centering}
\includegraphics[width=0.46\textwidth]{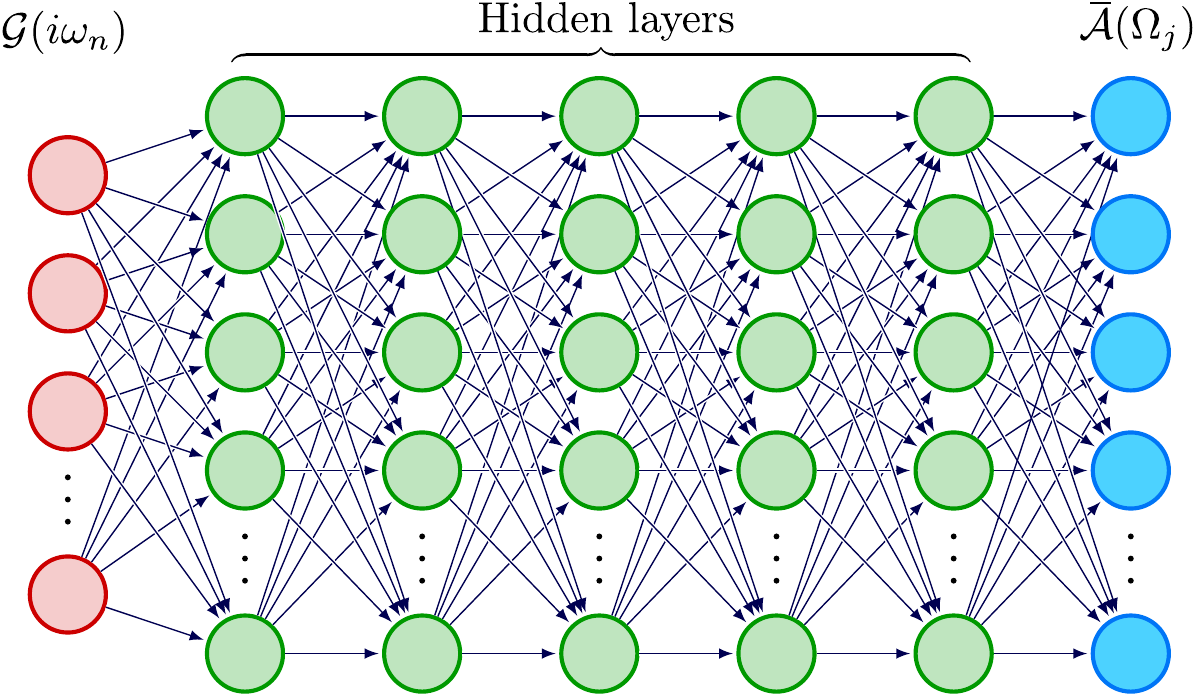}
\caption{Schematic of the fully-connected NN for the analytic continuation problem. The first layer (red) denotes the input of discretized imaginary-time Green's function ${\rm Re}/{\rm Im}[\mathcal{G}(i\omega_n)]$. The last layer (blue) denotes the output of discretized spectral function $\bar{\mathcal{A}}(\Omega_j)$. In between, there are five hidden layers (green) with the same number of neurons as the output layer.}
\label{FNN}
\end{centering}
\end{figure}

\section{Framework} 

Below we first introduce our general framework that includes both the training data preparation and the training scheme. 

To prepare training data, we need to include a variety of spectral functions $\mathcal{A}(\Omega)$. For each given type of $\mathcal{A}(\Omega)$, the corresponding imaginary-frequency Green's function $\mathcal{G}(i\omega_n)$ can be generated according to Eq. \ref{Analytical_continuation}. Each pair of $\{\mathcal{G}(i\omega_n), \mathcal{A}(\Omega)\}$ contributes one sample in the dataset. Since both the input and the output of NN are discrete, we shall first discretize $\mathcal{G}(i\omega_n)$ and $\mathcal{A}(\Omega)$. We discretize $\Omega$ into $N_\text{out}$ number of points $\Omega_j,\{j=1,\dots,N_\text{out}\}$ in a proper range of $\Omega$, and we normalize $\mathcal{A}(\Omega_{ {j}})$ to $\bar{\mathcal{A}}(\Omega_j)$ such that $\sum_{j=1}^{N_\text{out}}\bar{\mathcal{A}}(\Omega_j)=1$. Then, the output spectral function $\bar{\mathcal{A}}(\Omega)$ is represented by a discretized vector $\{\bar{\mathcal{A}}(\Omega_j){|}\{j=1,\dots,N_\text{out}\}\}$. For the input Green's function, $\omega_n=(2n+1)\pi/\beta$ where $n=1,\dots, N_\text{in}$ and $\beta$ is set as unity. $\mathcal{G}(i\omega_n)$ is generally a complex number. To mimic the inevitable noise in the calculated $\mathcal{G}(i\omega_n)$ from various numerical methods, we add a randomly sampled noise individually for each point $\omega_n$, i.e. 
\begin{eqnarray}
\tilde{\mathcal{G}}(i\omega_n)=\mathcal{G}(i\omega_n)(1+\delta+i\delta^\prime),\\
\end{eqnarray}
where noise $\delta$ and $\delta^\prime$ satisfy a Gaussian distribution with zero mean and standard deviation $\eta$. The input Green's function $\mathcal{G}(i\omega_n)$ is then represented by a discretized vector $\{\text{Re}{[}\tilde{\mathcal{G}}(i\omega_n){]},\text{Im}{[}\tilde{\mathcal{G}}(i\omega_n) {]} {|} \{n=1,\dots,N_\text{in}\}\}$.

The multi-layer fully connected NN used in this work is depicted in Fig.~\ref{FNN}. The input layer is the discretized $\mathcal{G}(i\omega_n)$, and the output layer is the discretized $\mathcal{A}(\Omega)$. The training process is carried out with the Adam optimizer \cite{Goodfellow}, and the Kullback-Leibler divergence (KLD) is used as the training loss function, which is defined as 
\begin{equation}
{\rm KLD} = \sum_{t}\sum_{j} \bar{\mathcal{A}}^t(\Omega_j)\log [\bar{\mathcal{A}}^t(\Omega_j)/\bar{\mathcal{A}}^t_{\rm NN}(\Omega_j)],
\end{equation}
where $t$ labels samples in the dataset. $\{\bar{\mathcal{A}}(\Omega_j)\}$ are the labels and $\{\bar{\mathcal{A}}_\text{NN}(\Omega_j)\}$ are the outputs of the NN.
After training, the mean absolute error on the testing set is chosen to evaluate the performance, which is defined as
\begin{equation}
\mathcal{L} =\frac{1}{\mathcal{N}} \sum_{t,j}\left| \bar{\mathcal{A}}^t_{\rm NN}(\Omega_j)-\bar{\mathcal{A}}^t(\Omega_j) \right|.
\end{equation}
Here the summation runs over all samples in the testing dataset, and $\mathcal{N}$ is the product of the dimension of the output and the number of samples in the testing dataset.
For all training processes, $10^5$ number of training samples and $5\times 10^3$ testing samples are prepared. 

\begin{figure}[t!!]
\begin{centering}
\includegraphics[width=0.42\textwidth]{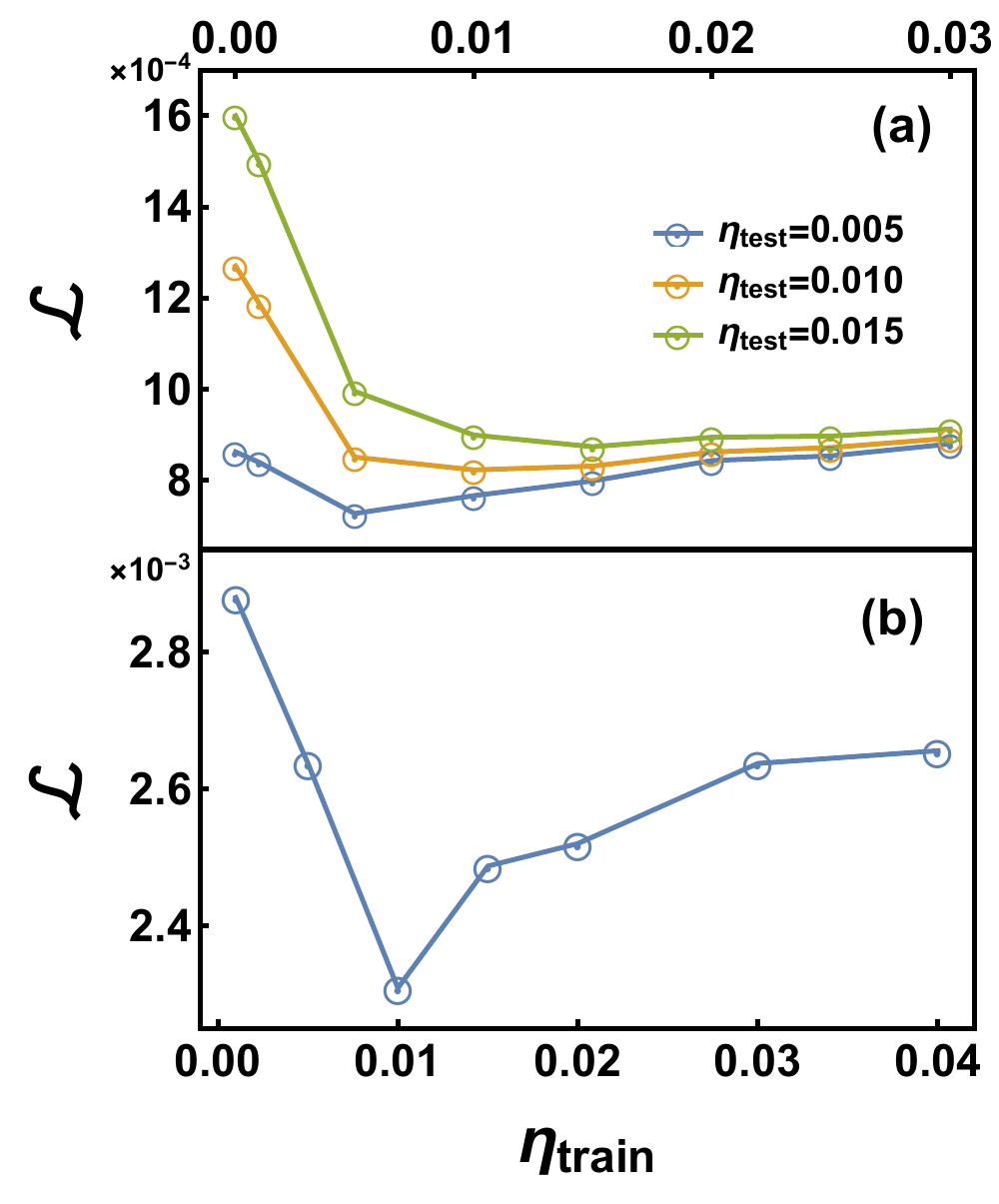}
\caption{The performance is measured by the mean absolute error $\mathcal{L}$ when the trained NN are applied to the testing dataset. The NN are trained with a given training dataset with noise strength $\eta_{\rm train}$, and $\mathcal{L}$ is plotted against $\eta_{\rm train}$. (a) Synthetic spectral function. Different curves correspond to different noise strength $\eta_{\rm test}$ of the testing dataset. (b) The transverse field Ising model. The quantum Monte Carlo simulation generates the testing dataset with unknown noise strength.  }
\label{performance}
\end{centering}
\end{figure}

\section{Results} 

Below we will present our results on two cases, one on synthetic spectral functions and the other on the transverse field Ising model. 

\subsection{Synthetic Spectral Functions} 

We generate the synthetic spectral function as a summation of Gaussian distributions.  
\begin{equation}
\mathcal{A}(\Omega)=\sum_{i}\lambda_i \mathcal{N}(\Omega | \mu_i,\sigma_i).
\end{equation}
Here $\mathcal{N}(\Omega | \mu,\sigma)$ denotes a Gaussian distribution centered at $\mu$ with width $\sigma$, that is 
\begin{equation}
\mathcal{N}(\Omega|\mu,\sigma)=\frac{1}{\sqrt{2\pi\sigma^2}}\exp\left\{ -\frac{1}{2\sigma^2}(\Omega-\mu)^2\right\}.
\end{equation}
The coefficients $\lambda_i$ take random value in $[0,1]$ under the constraint $\sum_i \lambda_i=1$, which ensures the normalization condition $\int d\Omega \mathcal{A}(\Omega)=1$. Here the maximum number of Gaussian components is three. The mean value $\mu$ of each Gaussian distribution ranges from $[-6,6]$ and the width $\sigma$ ranges from $[0.1,4]$. Since these synthetic spectral functions are made of Gaussian components, their corresponding imaginary-frequency Green's function $\mathcal{G}(i\omega_i)$ can be exactly calculated according to Eq. \ref{Analytical_continuation}.

\begin{figure}[t!!]
\begin{centering}
\includegraphics[width=0.42\textwidth]{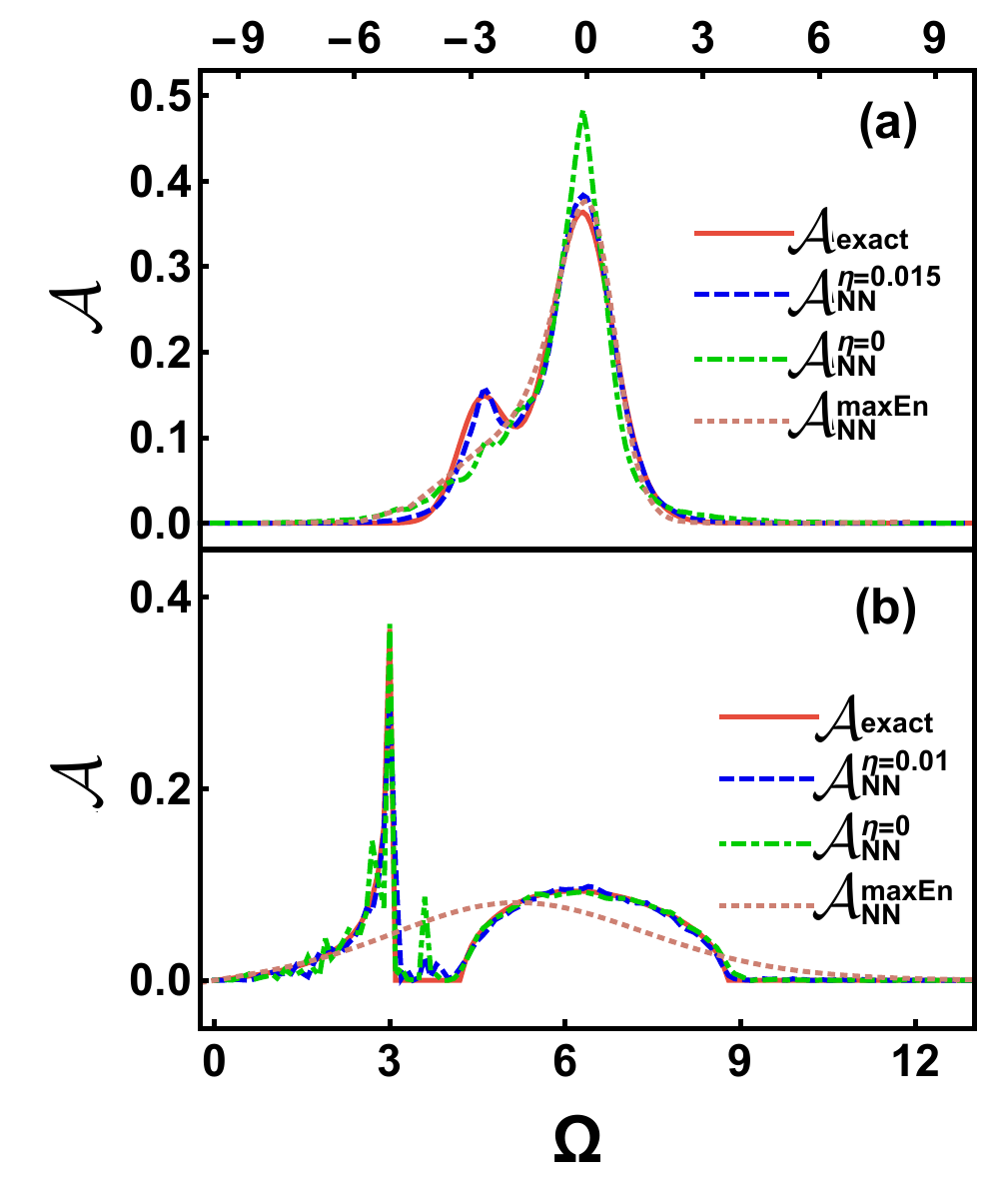}
\caption{Predictions of the NN. Red solid lines are the exact spectral functions. Blue dashed lines are the result of NN trained with the dataset with optimal noise. Green dot-dashed lines are the result of NN trained without noise. The brown dotted lines are the result of the maximum entropy method. (a) shows an example from synthetic spectral functions, and (b) shows an example from the transverse field Ising model. }
\label{comparing}
\end{centering}
\end{figure}

After training, the performance of the NN is measured by the mean absolute value $\mathcal{L}$ on the testing dataset. Here we add noise with amplitude $\eta_{\rm test}$ into the testing dataset in order to mimic the inevitable noise in numerical simulations. $\mathcal{L}$ is plotted in Fig.~\ref{performance}(a) for different values of $\eta_{\rm test}$.
Smaller of $\mathcal{L}$ indicates better performance of the trained NN. Fig. \ref{performance}(a) shows that, if $\eta_{\rm train}$ is below $\eta_{\rm test}$, $\mathcal{L}$ becomes large and the NN does not perform well. $\mathcal{L}$ approaches the optimal performance when $\eta_{\rm train} \approx \eta_{\rm test}$. Then, $\mathcal{L}$ slowly increases when $\eta_{\rm train} > \eta_{\rm test}$.

Fig. \ref{comparing} (a) shows a concrete typical example in the testing dataset with noise amplitude $\eta_{\rm test}=0.015$ . Here we show the NN results with and without noise in the training data, and we also show results from the MaxEnt method for comparison. As one can see, both the MaxEnt method (brown dotted line) and the NN results without noise (green dot-dashed line) miss the first peak. The MaxEnt method tends to generate a smooth and simple distribution with minimal singularities. In contrast, the results can be significantly improved when noise is added into the training data and when the noise amplitude is comparable with that in the test data. The blue dashed line shows that the NN result with $\eta_{\rm train}=\eta_{\rm test}=0.015$ can successfully capture both positions, amplitudes, and widths of two peaks in the spectral function, with tiny error compared with the exact results (solid red line). Hence, we show that our NN can successfully predict the spectral function, and the performance of the NN can be enhanced by adding noise properly into the training dataset. 

\subsection{The Transverse Field Ising Model} 

For synthetic spectral functions, we know precisely the amount of noise we add into the testing dataset. However, noise is always unavoidably generated from numerical simulations and its amplitude is generally not precisely known. To address such situations, we consider the one-dimensional transverse field Ising model, whose Hamiltonian reads 
\begin{equation} \label{HIsing}
	\hat{H} = -J\sum_{l}\hat{\sigma}_l^z\hat{\sigma}^z_{l+1}-g\sum_l\hat{\sigma}_l^x \, ,
\end{equation} 
where $\hat{\bm{\sigma}}_{l}$ are the standard Pauli operators on site $l$. Here $g$ is the transverse field strength, $J$ is the coupling along $z$-direction which will be set as a unit below, and the boundary condition is taken to be periodic.
Since this model can be solved exactly by mapping to free fermions using the Jordan--Wigner transformation, closed analytic forms of both $\mathcal{G}(i\omega_n)$ and $\mathcal{A}(\Omega)$ can be obtained (see Appendix A for details). 
Because the system is translational invariant, the calculation is mostly easily done in momentum space. Nevertheless, to keep our notation clean and consist, the dependences of physical quantities on the momentum quantum number $q$ and the parameter $g$ have been suppressed. In this way, a different choice of $q$ and $g$ generates a separate set of $\{\mathcal{G}(i\omega_n),\mathcal{A}(\Omega)\}$, which contributes one sample to the training dataset. For the testing dataset, the imaginary-time Green's function $\mathcal{G}(i\omega_n)$ is computed using a Worm-type quantum Monte Carlo method (see Appendix B for details) \cite{Huang:2020WormQIsing}, where noise with unknown strength is intrinsically embedded. 
As labels, their corresponding $\mathcal{A}(\Omega)$ are still obtained by the exact mapping to free fermions under the same parameters $q$ and $g$.

Fig. \ref{performance}(b) shows $\mathcal{L}$ as a function of $\eta_\text{train}$. In this case, it also shows a clear minimum around $\eta_\text{train}\approx 0.01$. This result means that the most proper noise strength also exists, even though the intrinsic numerical noise in the testing dataset may not be Gaussian type. With the optimum noise strength, the trained NN outputs the best prediction for the testing dataset produced by the quantum Monte Carlo simulations. One example is shown in Fig. \ref{comparing}(b). Compared with the synthetic spectral function shown in Fig.~\ref{comparing}(a), the spectral function of the transverse field Ising model is much more singular and contains sharp edges. Moreover, a dark continuum \cite{Goulko:2016DarkContinum}, a finite region with vanishing $\mathcal{A}(\Omega)$, exists between the two peaks. With the optimal noise strength, the trained NN correctly resolves both the positions and height of the two peaks, the locations of the sharp edges and the dark continuum. In comparison, the MaxEnt method fails to capture these features of the spectral function. The NN trained without noise also produces two peaks, but the fluctuations are much stronger, especially in the dark continuum. 

\section{Summary and Outlook} In summary, we have developed a NN-based method for analytic continuation. The training data is generated either from synthetic Gaussian type spectral functions or by the exact solution of the transverse field Ising model, where the analytic continuation can be done analytically. The validity of this method is demonstrated with data obtained from quantum Monte Carlo simulations. The main finding of this work is that a proper amount of noise has to be added to the training data in order to reach optimal performance. Our NN-based method can be easily applied to perform analytic continuation of Monte Carlo results from other models, such as the Bose-Hubbard model and the anisotropic Heisenberg models, and compare the analytic continuation results with experimental measurements of real-time dynamics, such as quantum simulation experiments with ultracold atomic gases.

\textit{Acknowledgment.} We thank Lei Wang for helpful discussions. This work is supported by NSFC Grant No. 11904190 (JY), 11734010 (HZ), and Beijing Outstanding Young Scientist Program (HZ).

\appendix
\section{Exact Solution for the Transverse Ising Model}
For the transverse Ising model, it is possible to obtain the analytic expressions of both the imaginary-time Green's function and the spectral function. The first step is to rotate the spins along the $y$-axis for negative $90$ degrees which defines the following new spin operators $\hat{\bm{\pi}}_{l}$, 
\begin{equation}
	\hat{\pi}_{l}^{x} = \hat{\sigma}_{l}^z \, , \quad \hat{\pi}_{l}^{y} = \hat{\sigma}_{l}^{y} \, , \quad \hat{\pi}_{l}^{z} = - \hat{\sigma}_{l}^{x} \, . 
\end{equation} 
The next step is to perform the standard Jordan--Winger transformation that maps spin operators to fermion operators $\hat{a}_l^{\dagger}$ and $\hat{a}_l$
\begin{equation}
\left\{\begin{aligned}
\hat{\pi}^+_l&=(\hat{\pi}^x_l+i\hat{\pi}_l^y)/2=(-1)^{\sum_{m<l}\hat{n}_m}\hat{a}_l^\dagger \, , \\
\hat{\pi}^-_l&=(\hat{\pi}^x_l-i\hat{\pi}_l^y)/2=(-1)^{\sum_{m<l}\hat{n}_m}\hat{a}_l 
, , \\
\hat{\pi}^z_l&=2\hat{a}_l^\dagger \hat{a}_l-1 \, ,
\end{aligned}\right.
\end{equation}
where $\hat{n}_{m} = \hat{a}^{\dagger}_{m} \hat{a}_{m}$ is the fermion occupation operator at site $m$. In terms of these fermionic operators, the resulting Hamiltonian of Eq.~(\ref{HIsing}) is quadratic, 
\begin{equation} \label{eq:H_a_fermion}
	\hat{H}=-J\sum_l(\hat{a}_l^\dagger-\hat{a}_l)(\hat{a}_{l+1}+\hat{a}_{l+1}^\dagger)+g\sum_l(2\hat{n}_l-1) \, . 
\end{equation}
The boundary condition becomes $\hat{a}_{L+1}^{\dagger} = -\hat{\eta} \hat{a}_{1}^{\dagger}$ where $L$ is the number of sites and $\hat{\eta} = (-1)^{\sum_{l} \hat{n}_{l}}$ is the fermion parity operator. After a phase rotation of the operator $\hat{a}_{l}$ by $\pi/4$, in momentum space the Hamiltonian assumes the form 
\begin{equation}
\begin{aligned}
&\hat{H}= \\
&\sum^\prime_k
\begin{bmatrix}
\hat{a}^\dagger_k & \hat{a}_{-k}
\end{bmatrix}
\begin{bmatrix}
2g-2J\cos(kd) & 2J \sin(kd) \\
2J \sin(kd) & 2J\cos(kd)-2g
\end{bmatrix}
\begin{bmatrix}
\hat{a}_k \\ \hat{a}^\dagger_{-k}
\end{bmatrix}
\end{aligned}
\end{equation}
where the prime means summation over different pairs of $k$ and $-k$ except for $k=0$, $d$ is the lattice spacing, and $\hat{a}_k=e^{i\frac{\pi}{4}}\frac{1}{\sqrt{L}}\sum_{l=1}^{L} e^{ilkd}\hat{a}_l$ with $L$ being the lattice size. One can then easily diagonalize the Hamiltonian by the standard Bogoliubov transformation, resulting in a free fermion Hamiltonian
\begin{equation} \label{eq:H_f_fermion}
	\hat{H} = \sum_k^\prime \xi_k (\hat{f}_k^\dagger\hat{f}_k +\hat{f}_{-k}^\dagger\hat{f}_{-k} ), 
\end{equation}
where $\hat{f}_{k}^{\dagger}$ ($\hat{f}_{k}$) is the new fermion creation (annihilation) operator of momentum mode $k$ and $\xi_k = 2\sqrt{J^2+g^2-2gJ\cos(kd)}$ is the corresponding quasi-particle energy. The explicit expression of the Bogoliubov transformation reads 
\begin{equation}
\begin{bmatrix}
\hat{a}_k \\ \hat{a}^\dagger_{-k}
\end{bmatrix} =
\begin{bmatrix}
\cos\theta_k & -\sin\theta_k \\
\sin\theta_k & \cos\theta_k
\end{bmatrix}
\begin{bmatrix}
\hat{f}_k \\ \hat{f}^\dagger_{-k}
\end{bmatrix}
\end{equation}
where $\theta_k$ is determined by $\tan2\theta_k = J\sin(kd)/[g-J\cos(kd)]$.

We choose to study the imaginary-time spin-spin correlation function along the $x$-direction,
\begin{equation} \label{eq:Gxx}
\mathcal{G}(l,\tau)=-\frac{1}{4}\langle \mathcal{T}_{\tau} \hat{\sigma}_l^x(\tau)\hat{\sigma}^x_0(0)\rangle,
\end{equation}
where $\hat{\sigma}_l^x(\tau)=e^{\tau\hat{H}}\hat{\sigma}_l^xe^{-\tau\hat{H}}$ is the imaginary-time Heisenberg operator and $\mathcal{T}$ denotes the time-ordering operator. This correlator corresponds to the density-density correlation function in terms of the Jordan--Wigner fermion $\hat{a}_{l}$. Since the Hamiltonian Eq.~\eqref{eq:H_a_fermion} or \eqref{eq:H_f_fermion} is quadratic, the correlator can be evaluated using Wick’s theorem. 
After a tedious but straightforward calculation, the momentum and frequency representation of the correlator $\mathcal{G}(q,i\omega_n)$,
\begin{equation}
\mathcal{G}(q,i\omega_n) =\frac{1}{L} \sum_{l} e^{-iqld}  \int_{0}^{\beta} \mathcal{G}(l,\tau) \, \mathrm{e}^{i\omega_{n} \tau }d\tau,
\end{equation}
which takes the following explicit form
\begin{equation}
\begin{aligned}
&\mathcal{G}(q\neq 0,i\omega_n)=\\
 &\sum_{k} 
(f_{k+q}+ f_{k}-1)\sin(\theta_{k+q}+\theta_{k})\times  \\
&\frac{ i\omega_n \sin(\theta_k-\theta_{k+q})+(\xi_{k+q}+\xi_k)\sin(\theta_k+\theta_{k+q})}{\omega_n^2+(\xi_{k+q}+\xi_k)^2} \\
+ &\sum_{k}(f_{k+q}- f_{k})\cos(\theta_{k+q}+\theta_{k})\times  \\
&\frac{ i\omega_n \cos(\theta_k-\theta_{k+q})+(\xi_{k+q}-\xi_k)\cos(\theta_k+\theta_{k+q})}{\omega_n^2+(\xi_{k+q}-\xi_k)^2} .
\end{aligned}
\label{EqG}
\end{equation}
Here $f_k$ is the Fermi-Dirac distribution function $f_k=1/(e^{\beta\xi_k}+1)$ and $\omega_n =2\pi n/\beta$ is the even Matsubara frequency where $n=1,\dots, N_\text{in}$. With the analytical formula of $\mathcal{G}(q\neq 0,i\omega_n)$,
we can then analytically perform the analytic continuation and read out the spectral function,
\begin{equation}
\begin{aligned}
& \mathcal{A}(q \neq 0, \Omega) = -\frac{1}{\pi} \operatorname{Im} \mathcal{G}(q\neq 0, i\omega_n \to \Omega+i 0^+) \\
= & \sum_k    (1-f_{k+q}- f_{k})\times \\
[& \sin(\theta_{k+q}+\theta_{k}) \cos\theta_{k+q}\sin\theta_{k}\delta(\Omega-\xi_k-\xi_{k+q})- \\
&\sin(\theta_{k+q}+\theta_{k})\sin\theta_{k+q}\cos\theta_k\delta(\Omega+\xi_k+\xi_{k+q})] \\
+&\sum_k (f_{k}- f_{k+q})\times \\
[&\cos(\theta_{k+q}+\theta_{k})\cos\theta_{k+q}\cos\theta_{k}\delta(\Omega+\xi_k-\xi_{k+q})+\\
&\cos(\theta_{k+q}+\theta_{k})\sin\theta_{k+q}\sin\theta_k\delta(\Omega-\xi_k+\xi_{k+q})] .
\end{aligned}
\label{EqA}
\end{equation}
In conclusion, both the Green's function or the correlator $\mathcal{G}(q,i\omega_n)$ and the spectral function $\mathcal{A}(q,\Omega)$ can be evaluated exactly according to Eq. \eqref{EqG} and \eqref{EqA} for each $g$, with $J$ being the energy unit. For preparation of the training dataset, setting temperature $\beta=2.0/J$ and interaction strength $g$ ranging from $0.5J$ to $2.0J$ with interval $0.1J$.   For each parameter $g$, the momentum $q$ will be discretized into $(0, 0.01, 2\pi]$ and setting $d$ as the unit length.Here in this transverse problem, $\mathcal{A}$ doesnot obey the normalization relation $\int d\Omega \mathcal{A}(\Omega)=1$. Thus the mean-square error is adopted for the training process.

\section{Quantum Monte Carlo Simulation of the Transverse Ising Model}
We closely follow Ref. \cite{Huang:2020WormQIsing} to map spin operators into hard-core boson operators such that the resulting partition function can be efficiently sampled by the Worm algorithm \cite{Prokofev:1998WormJETP,Prokofev:1998WormPhysLettA}. We first perform a spin rotation around the $y$-axis for $90$ degrees 
and the spin rotational operators $\hat{\pi}_{l}$ is defined as
\begin{equation}
   \hat{\pi}_{l}^{x} = - \hat{\sigma}_{l}^z \, , \quad \hat{\pi}_{l}^{y} = \hat{\sigma}_{l}^{y} \, , \quad \hat{\pi}_{l}^{z} = \hat{\sigma}_{l}^{x} \, . 
\end{equation}
For the negative rotation, the corresponding boson Hamiltonian in this time can be written as a sum of the kinetic term $\hat{K}$ and the potential energy term $\hat{U}$, 
\begin{equation}
    \hat{H} = \hat{K} + \hat{U} \, ,
\end{equation}
where \begin{align}
  \hat{K} & \equiv \hat{K}_{1} + \hat{K}_{2} \\
          & \equiv -J \sum_{l}(\hat{a}_{l}^{\dagger} \hat{a}_{l+1}+\text {h.c.})- J \sum_{l} (\hat{a}_{l}^{\dagger} \hat{a}_{l+1}^{\dagger}+\text {h.c.})
\end{align}
and 
\begin{equation}
    \hat{U} = -2 g \sum_{l} \hat{a}_{l}^{\dagger} \hat{a}_{l}  \, .
\end{equation}
Here we have omitted unimportant constants. Employing the expansion 
\begin{widetext}
     $ \displaystyle e^{-\beta \hat{H}} = e^{-\beta \hat{U}} \mathcal{T}_{\tau} \exp \left[-\int_{0}^{\beta} \hat{K}(\tau) d \tau\right] = e^{-\beta \hat{U}} \sum_{n=0}^{\infty} (-1)^{n} \int_{0}^{\beta} d \tau_{n} \cdots \int_{0}^{\tau_{2}} d \tau_{1} \left(\hat{K}_{1}\left(\tau_{n}\right)+\hat{K}_{2}\left(\tau_{n}\right)\right) \cdots\left(\hat{K}_{1}\left(\tau_{1}\right)+\hat{K}_{2}\left(\tau_{1}\right)\right) \, ,$ 
\end{widetext}
where $\mathcal{T}_{\tau}$ is the imaginary time ordering operator and $\hat{K}(\tau) = e^{\tau \hat{U}} \hat{K} e^{-\tau \hat{U}}$. In the Fock basis a typical term (out of $2^{n}$ terms) in the $n$-th order expansion of the partition function reads 
\begin{equation} \label{eq:conf_weight}
    J^{n} \left( \prod_{j=1}^{n} \int d \tau_{j} \right) \cdot \exp \left[-\int_{0}^{\beta} U(\tau) d \tau\right],
\end{equation}
since the matrix element of the both hopping and pair creation/annihilation is one. Here $U(\tau)$ is the potential energy of the hard-core bosons at imaginary time $\tau$. As is clear from Eq.~(\ref{eq:conf_weight}), the weight of the configuration is positive and can be sampled by the Monte Carlo method. Following the detailed update procedures in \cite{Huang:2020WormQIsing}, the partition function can be efficiently sampled by Worm-type algorithms. The imaginary-time Green's function $\mathcal{G}(l-l^\prime, \tau)$ defined in Eq.~\eqref{eq:Gxx} just corresponds to $-\langle \mathcal{T}_{\tau} \hat{n}_{l}(\tau) \hat{n}_{l^\prime}(0) \rangle$ asides from unimportant constants. Since $\hat{n}_{l}(\tau)$ is diagonal in the Fock basis, the statistics can be easily accumulated in Monte Carlo simulations. We perform the simulation with system size $L=100$ at various values of transverse field $g$ to prepare the training and testing datasets.

\end{document}